\documentclass[11pt]{article}
\RequirePackage{latexsym}
\RequirePackage{graphicx}
\RequirePackage{amssymb}
\RequirePackage{setspace}

\usepackage{amsmath}
\usepackage{txfonts}
\usepackage[font=normalsize]{caption}
\usepackage{hyperref}
\usepackage{float}

\usepackage{wrapfig}
\usepackage{graphicx}
\usepackage{subcaption}
\usepackage[round,semicolon,authoryear]{natbib}

\topmargin -0.8in\graphicspath{ {figures/} }
\usepackage{array}
\chardef\us=`\_

\usepackage{color}
\usepackage{array}
\usepackage{url}
\usepackage{ulem}
\usepackage{hyperref}
\usepackage{appendix}
\usepackage{wrapfig}
\usepackage{gensymb}
\usepackage{float}
\usepackage[dvipsnames]{xcolor}
\usepackage[font=small]{caption} 

\begin{document}


     \begin{figure}
     \vskip 0.5in
     \begin{center}
     \includegraphics[scale=0.15]{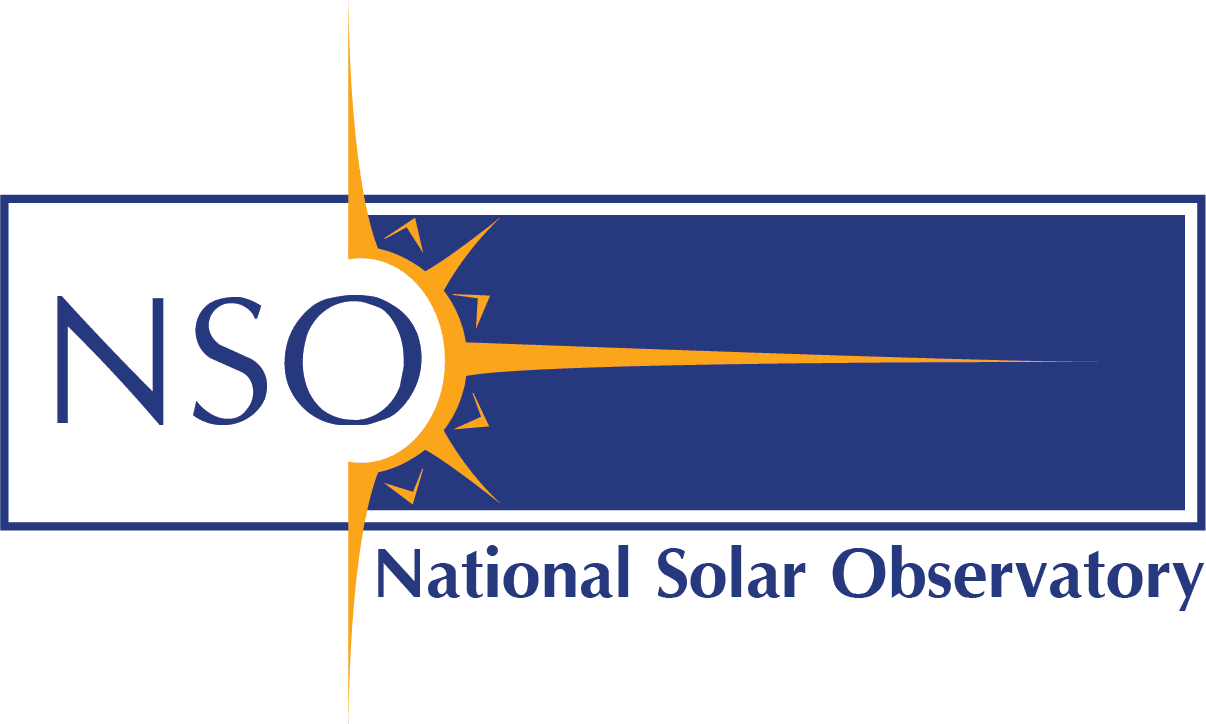}
     \end{center}
     \end{figure}

 \begin{center}
  \vskip 1in
{\bf\LARGE Improvements to the NSO Farside Mapping Pipeline: Noise Reduction Updates }

    \vskip 1in
{\Large  Mitchell Creelman, Kiran Jain, Niles Oien, John Britanik, \\ and Thomas M. Wentzel}\\
\vskip 0.5in
 {\it\large NISP, National Solar Observatory, Boulder, CO 80303, USA}
\end{center}

\begin{center}
Technical Report No. {NSO/NISP-2025-001}
\end{center}

\vskip 0.5in

\noindent\rule{\linewidth}{0.5mm}
\rule{\linewidth}{0.5mm}
    
    \begin{abstract}
    \addcontentsline{toc}{subsection}{\bf Abstract}
    
        The National Solar Observatory (NSO)'s Farside Pipeline is a critical tool of the space weather industry. It enables the detection and tracking of solar active regions that have rotated to the farside (invisible surface) of the Sun without relying on direct observational platforms such as satellites.  By applying the technique of helioseismic holography to continuous Doppler images of the front side (visible surface), the pipeline infers the size and location of these regions through the acoustic signatures. These farside maps, produced using data from the NSO's GONG Network, allow scientists and solar observers to monitor the behavior of solar active regions. They support efforts to protect vital telecommunications and national interest infrastructure. While the data from this pipeline are widely used for many scientific, industrial, and national security applications, global helioseismic monitoring remains a developing field, with ongoing refinements in methodology and reliability.  In this report, we will outline the updates made to the NSO's Farside Pipeline which have resulted in more accurate and consistent helioseismic maps, strengthening its value for both operational forecasting and scientific research.    
    \end{abstract}

\pagebreak

    \tableofcontents

\pagebreak

\pagebreak
\section[{Introduction}]{{Introduction}}
    \label{sec: Introduction}

\begin{figure} [h]
            \begin{center}
              \includegraphics[width=0.7\textwidth]{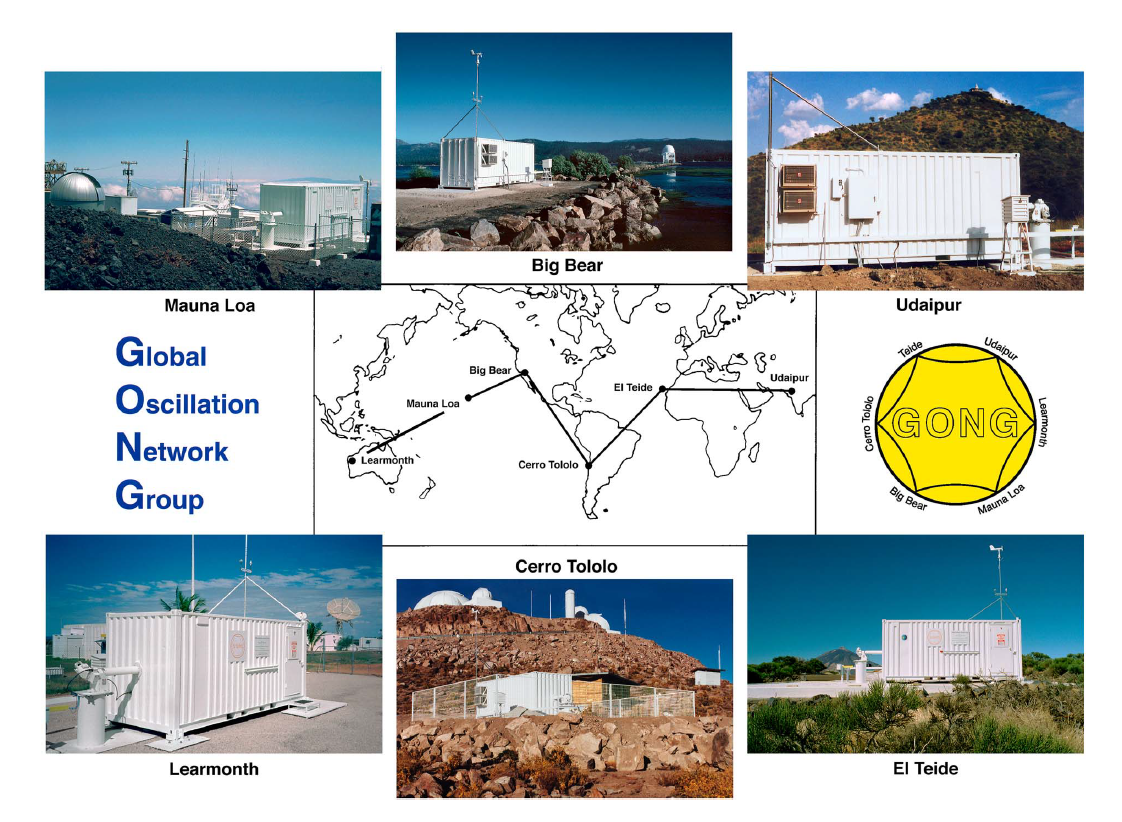}
                \caption{\it Global distribution of six sites of the  GONG network.}
                \label{fig: sites}
            \end{center}
        \end{figure}

    The National Science Foundation's (NSF) Global Oscillation Network Group (GONG) is a global observatory network managed by the National Solar Observatory's (NSO) Integrated Synoptic Program (NISP), consisting of 6 observational sites around the globe. These sites are shown in  Figure~\ref{fig: sites}. GONG has been collecting continuous solar observations at a cadence of 1 min since 1995 \citep{Harveyetal1996,hill2018}. The selection of sites was carefully considered in order to maximize the number of continuous minutes when observations were being collected \citep[e.g., see][for details]{jainContinuousSolarObservations2021}. GONG provides images for several observables, including the Dopplergrams, which are the main focus in this report.  \\

    Dopplergrams collected at individual sites across the GONG network are  analyzed for various helioseismic studies. This report primarily employs Dopplergrams through the application of helioseismic holography, a local helioseismology technique. This method provides the foundation for mapping active regions on the Sun’s farside \citep{lindseySeismicImagingSuns2017}. These maps provide a critical role for space-weather forecasters, enabling the continuous tracking of the movement and behavior of solar active regions while they are not directly visible from the Earth's surface \citep{jainSeismicMonitoringSuns2023}. Helioseismic holography has played an important role in providing valuable insights for space-weather forecasting; however, the quality of seismic maps can be affected by various factors, leading to potential noise. The work in this report outlines the most recent steps we have taken to significantly reduce the inherent noise in farside active region maps and improve the presentation of the maps. 

\section[{Background}]{Background}
    \label{sec: Background}

    The foundational element in both the legacy pipeline (currently operational) and the updated version (outlined in this report and slated to replace the legacy system) is the 24-hour phase shift map. These maps are generated from Dopplergrams collected over a full day at each GONG network site. In the farside pipeline, we do not use full-resolution Dopplergrams, instead these are binned down to 215 $\times$ 215 pixels at individual sites and transferred to GONG HQ in Boulder. These reduced-resolution Dopplergrams are labeled with the filename {\tt fqi}. For every minute, high-quality simultaneously observed site images are merged into a single Dopplergram—resulting in 1440 Dopplergrams per day. 
    These are then processed to create a phase shift map using the helioseismic technique mentioned in the previous section. The ratio of minutes throughout the day where at least one site in the GONG network collected observations is then recorded as the duty cycle. As discussed in \citet{jainSeismicMonitoringSuns2023}, the duty cycle $\ge$ 80\% is required for a phase shift map to be of adequate quality. 

    \subsection[{Farside Pipeline Structure}]{Farside Pipeline Structure}
        \label{sec: f6u Pipeline Overview}

\begin{figure} [h]
            \begin{center}
              \includegraphics[width=0.75\textwidth]{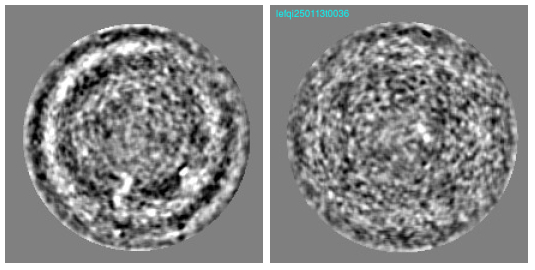}
                \caption{A contaminated "dinner plate"(left) and an uncontaminated  (right) {\tt fqi} Dopplergrams.}
                \label{Example FQI}
            \end{center}
        \end{figure}

        The input data for NSO farside pipeline is the {\tt fqi} Dopplergrams. This product is generated by the quick reduce software routine at the GONG sites. The {\tt mrfqi} product is a Dopplergram that is generated at a one minute cadence. Sometimes,  artifacts may appear in {\tt <site>fqi} Dopplergrams. This is often the case for Dopplergrams collected immediately after a calibration sequence is performed at the GONG site. These artifacts are informally referred to as “dinner plate” images, named for their visual resemblance to a dinner plate caused by oscillatory distortions near the solar limb (e.g., Figure~\ref{Example FQI}). Additional types of artifacts may also arise from instrument malfunctions, minor glitches, or suboptimal observing conditions. As part of this work, a machine learning algorithm was developed to remove contaminated images such as these \citep[see ][for details]{creelmanAnomalyDetectionGONG2024}.

        These site Dopplergrams are sent to the NISP data center in near-real time where they are merged by averaging them for a given minute. One may expect a continuous sequence of merged Dopplergrams in ideal conditions. In practice, observational gaps occur during periods when no site within the network successfully acquires data. As a results, this introduces gaps in the time series.\\

        It is worth mentioning that the concept of selecting the optimal site Dopplergram for each individual minute, rather than merging all available site Dopplergrams for that minute, has been considered. However, implementing this approach would require a robust and objective criterion for identifying the “best” Dopplergram at any given time. At present, the Dopplergrams are simply merged. 

\begin{figure} [h]
            \begin{center}
           \includegraphics[width=0.41\textwidth]{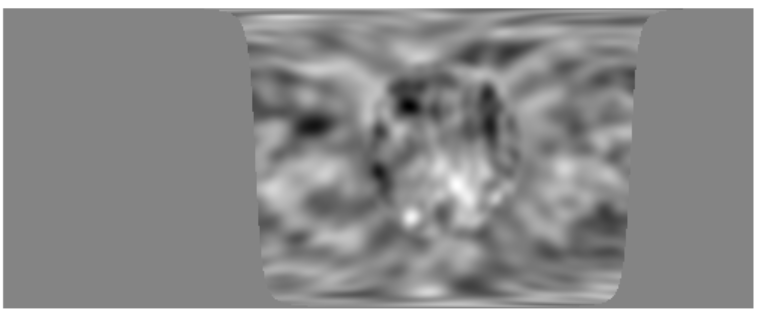}
           \includegraphics[width=0.41\textwidth]{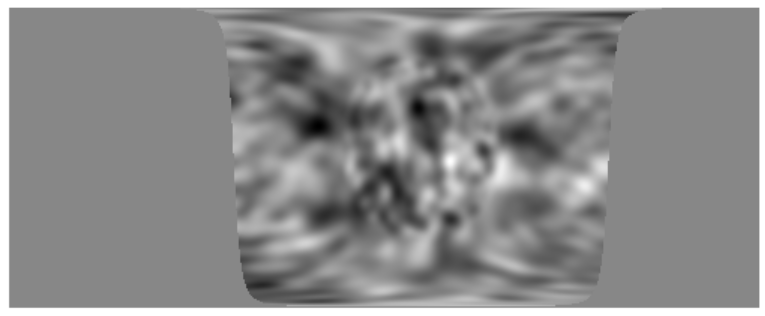}
        \includegraphics[width=0.41\textwidth]{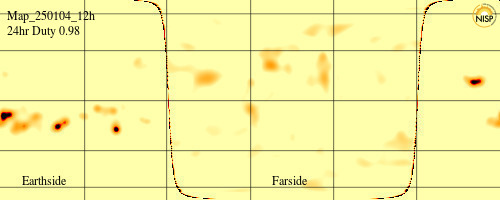}
         \includegraphics[width=0.41\textwidth]{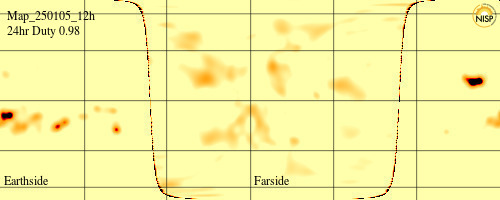}
                \caption{(Top row) Phase-shift maps of the farside for timestamps 20250104t1200 (left) and 20250105t1200 (right) generated by the legacy pipeline, classified as an fqm products. (Bottom row) Regions of higher negative phase shifts extracted from the {\tt fqm} maps are shown in the top row, an {\tt fqj} product. The images
                    also show the scaled magnetogram on the Earth-side.
                    The demarcation between the two sides is marked by nearly vertical lines.
                    Since {\tt fqj} product combines phase shifts with observed magnetic field, it is not entirely consistent over the solar sphere. As such, it is primarily intended for symbolic representation rather than quantitative analysis.}
                \label{Phase_Map}
            \end{center}
        \end{figure}

        As mentioned above, the core of the farside pipeline is the generation of 24-hr phase maps. The first step is the processing of site Dopplergrams ({\tt <site>fqi}). The images first undergo quality control checks before being merged and are used to generate a phase map of the hemisphere of the Sun that faces away from Earth (i.e., the farside). These phase maps (the {\tt mrfqm} product in the legacy pipeline) reflect phase shifts in wave travel that can be deduced from the input Doppler observations. A significantly negative phase shift (which corresponds to a wave arriving back at the front side of the Sun earlier than expected) has been found to be indicative of an active region at a corresponding point on the farside of the Sun. We display an example of the phase map in Figure~\ref{Phase_Map}.\\

        Though the analysis sketched above is rather involved and is not the primary focus of this report. It is covered briefly here. It is worth noting that ideally, the process operates on an uninterrupted series of Dopplergrams. Interruptions due to visual obstructions, instrumentation issues, or outages at the GONG sites can introduce noise into the output phase map. It is also worth noting that the phase shifts are quantatively small, and thus distinguishing signals within the noise can prove difficult. When expressed in radians with a negative phase indicating a wave returning earlier than expected, a phase shift more negative than about $-$0.1 radians is generally indicative of an active region on the farside, assuming the phase map is not excessively noisy.
        
\begin{figure} [t]
            \begin{center}
              \includegraphics[width=0.69\textwidth]{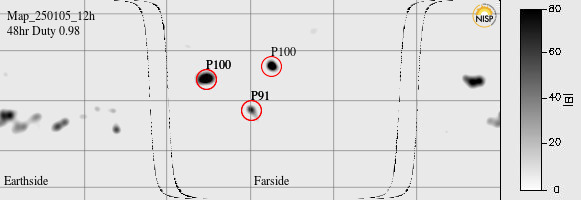}
                \caption{A legacy pipeline product, {\tt fqj},  that combines observed magnetic field 
                on the frontside with farside pipeline estimated magnetic field strength. The farside map is created by combining two {\tt fqm} maps. Since the mean duty cycle is 98\%, the sequence of input Dopplergrams was largely uninterrupted and the map is not noisy.
        Three candidate active regions have been identified, marked with red circles 
        and labeled with the probability for their appearance on the front side based 
        on the estimated strength. These labels are only applied if the duty cycle is
        high enough for the output map to be considered trustworthy. The edges of the
        two {\tt fqm} maps are marked by nearly vertical lines. As the map translates phase shifts into calibrated magnetic field strengths, the resulting product is commonly referred to as the farside calibrated map.}
                \label{Mag_fqg}
            \end{center}
        \end{figure}

        The estimates of the magnetic strength of farside active regions have been derived by analyzing  phase shift values of the farside active regions  and frontside magnetic field strength of a given region once it comes to the front side, where it can be directly observed \citep{igh_2007}. This estimate is used in the final calibrated farside map and to generate the fqg product for {\tt fqg} mqaps,  we combine two consecutive {\tt fqm} maps separated by 24 hr which minimizes the random noise.  These maps are also referred to as {\it composite} maps. Figure~\ref{Mag_fqg} displays an example of {\tt fqg} map where two maps shown in the top row of Figure~\ref{Phase_Map} are combined and converted to magnetic field map using a relation between the phase shifts and magnetic field strength. The map is composed of two parts, the farside estimated magnetic field, and the frontside observed magnetic field. This allows continuous monitoring of the solar magnetic activity in near real time with direct and indirect observations, a capability with has proven extremely useful for space weather modelers.

        The next section outlines the major steps used to construct 24-hour phase-shift maps.

    \subsection[{Legacy Composite Map Creation }]{Legacy Composite Map Creation}
        \label{sec: Legacy Composite Map Creation (fsWeb)}
        
    Composite maps are generated in two stages: first, individual 24-hour (component) maps are produced; then, two of these maps are combined to create the final composite map. The first step  follows a structured process in five stages. The stages are :

        \begin{enumerate}
        
            \item Re-project 1-min cadence {\tt fqi} Dopplergrams (i.e. 1440 for 24 hr) are onto the same central point to create a set of Postel-projected images. There is also an option to "collapse", or do averaging, on the Dopplergrams, although that is currently not done in production.
            
            \item Stack image into a data cube.  Transpose  this data cube so that the time index is $x$.  Analyze the temporal profile of each pixel removing its average, its linear trend, and setting pixels in the temporal sequence that are apparently bad to zero.  Fourier-transform the data in time and transpose the data cube back to the original configuration.
            
            \item Extract the specified frequency interval of 2.5 to 4.5 mHz in the spectrum and transpose the data cube.
            
            \item Compute wave ingression and egression as 3D complex images. This step is actually repeated three times since it is dependent on the skip pattern, that is, the number of reflections that take place inside to Sun on the wave's journey. This step is done for the skip patterns 3 $\times$ 1, 2 $\times$ 2, and 1 $\times$ 3 to cover the full hemisphere. and the results are then combined.
            
            \item Derive farside image from the correlation map of egression and ingression using three-dimensional complex images from the previous stage. It produces complex correlation map between egression and ingression, and finally full-hemisphere farside map in (Lon, Sine(Lat)) projection is created for 24 hr. 
            
       
        \end{enumerate}     

\section[{Updates and Improvements}]{Updates and Improvements}
    \label{sec: Updates and Improvements}
    
    We have identified multiple opportunities to enhance the product quality  within the farside pipeline. While all of these fall within the project's scope of increasing the image quality of farside maps produced by GONG, many of them were only discovered after the initiation of the project. The primary goal is to reconfigure composite image generation and introduce new visualizations and data products at the pipeline. However, we soon found that some individual site images were contributing substantial noise to the final data products. The current iteration of the updated pipeline includes an AI based filter to assure {\tt fqi} image quality from individual sites, an increased temporal resolution, variable averaging of composite maps or improved calibration near limbs, a duty cycle threshold for 24 hour phase shift maps,  and improved processing and visualization for the farside maps. These are discussed below in details.  Potential further improvements to the updated pipeline will be outlined later in this report.  
        
      \subsection[{Reducing Random Noise}]{Reducing Random Noise }
        \label{sec: Noise}  
        As mentioned earlier, one of the main goals of the National Science Foundation funded  Windows on the Universe (WoU) Project is to improve the  farside helioseismic monitoring products by reducing the inherent noise. This will  provide a far more reliable data product to the space weather community and reduce the need for expensive, and temperamental space-based observational platforms in space-weather monitoring. This section provides a detailed account of the steps we’ve taken to advance this goal.\\

       The first step is to improve the quality of the merged Dopplergrams. This requires immediate measures for implementing automated data quality checks for site Dopplergrams prior to merging \citep{creelmanAnomalyDetectionGONG2024}, excluding maps generated from low duty cycle observations, and incorporating additional 24 hour phase shift maps to enhance the 48-hour calibrated products. 
         Further and future planned improvements include the re-design of the 'find candidate' methodology for identifying and labeling active regions. As previously mentioned, the 6 hourly cadence maps are processed identically to those produced by the legacy pipeline, drawing on the same filtered site images. The differences, however, are in the processing steps these maps undergo before they are finally produced as 6 hour composites. The temporal cadence, averaging methods, magnetic strength calculations, and other aspects outlined further in this report, have all been updated. 

        \begin{wrapfigure}{r}{0.35\textwidth}
          \vspace{-10mm}
            \includegraphics[width=0.35\textwidth]{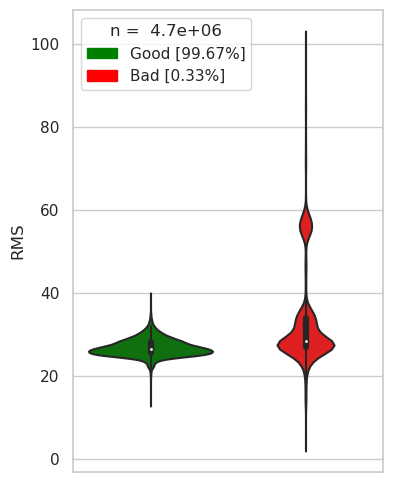}
                  \captionsetup{font=footnotesize}
                \caption{Violin plot showing distribution of network identified good and anomalous images with their corresponding RMS values. Note the frequent occurrence of overlapping rms distributions below 40.}
                \label{fig:RMS_Violin}
        \end{wrapfigure}
    
    \subsubsection[{Removing Erroneous Site Images: FQI Filter}]{Removing Erroneous Site Images: FQI Filter}
        \label{sec: FQI Filter}
        One of the primary sources of noise in the legacy farside pipeline was the inclusion of anomalous Dopplergrams from the individual sites.
        Anomalous images can stem from a wide variety of sources, including mechanical issues with the cameras, or software issues from the post-capture processing. The largest challenge for excluding anomalous images in this process is variety of statistical characteristics and the volume of the data. For a typical day, about  four thousand site images are collected by the GONG network. The large volume of data makes manual review in near-real time  a highly intensive process which would require a disproportionate amount  of time and manpower to accomplish. Previously, an RMS threshold of 60 was used to exclude  anomalous site images. While this did catch some, it failed to exclude many others from entering the data stream, as shown in Figure~\ref{fig:RMS_Violin}. \\
              
        In order to better identify and exclude anomalous images from the farside pipeline, a fully convolutional network was developed \citep[see,][for details]{creelmanAnomalyDetectionGONG2024,Jain2025SoPh}. This AI quality assessment filter was trained on hand-labeled site images and has proven highly effective at identifying and excluding anomalous data from the {\tt mrfqi} process. For more details on the  performance and design of this filter, please refer to  
        \citet{creelmanAnomalyDetectionGONG2024}.

    \subsubsection[{Increased Cadence of 24 hour phase shift maps}]{ Increased Cadence of 24 hour phase shift maps}
        \label{sec: Increased Cadence}
    
         In the legacy farside pipeline, composite phase shift maps are constructed from two independent 24-hour phase shift maps separated by a 24-hour interval. This results in 48 hours of total coverage in each composite farside phase shift map. While the temporal coverage for these maps is adequate, the low-temporal cadence does not provide much resilience against random noise in the final products. The solution to this has been to increase the temporal resolution for the composite maps, using a maximum of 5 maps generated at a 6 hour cadence. This allows for the suppression of noise in the final product and provides a larger number of 24 hour phase shift maps include in the final composites. 
         
        \begin{figure}[t]
        \centering
            \begin{subfigure}[b]{\textwidth}
                \hspace{7mm} 
                \includegraphics[width=0.79\textwidth]{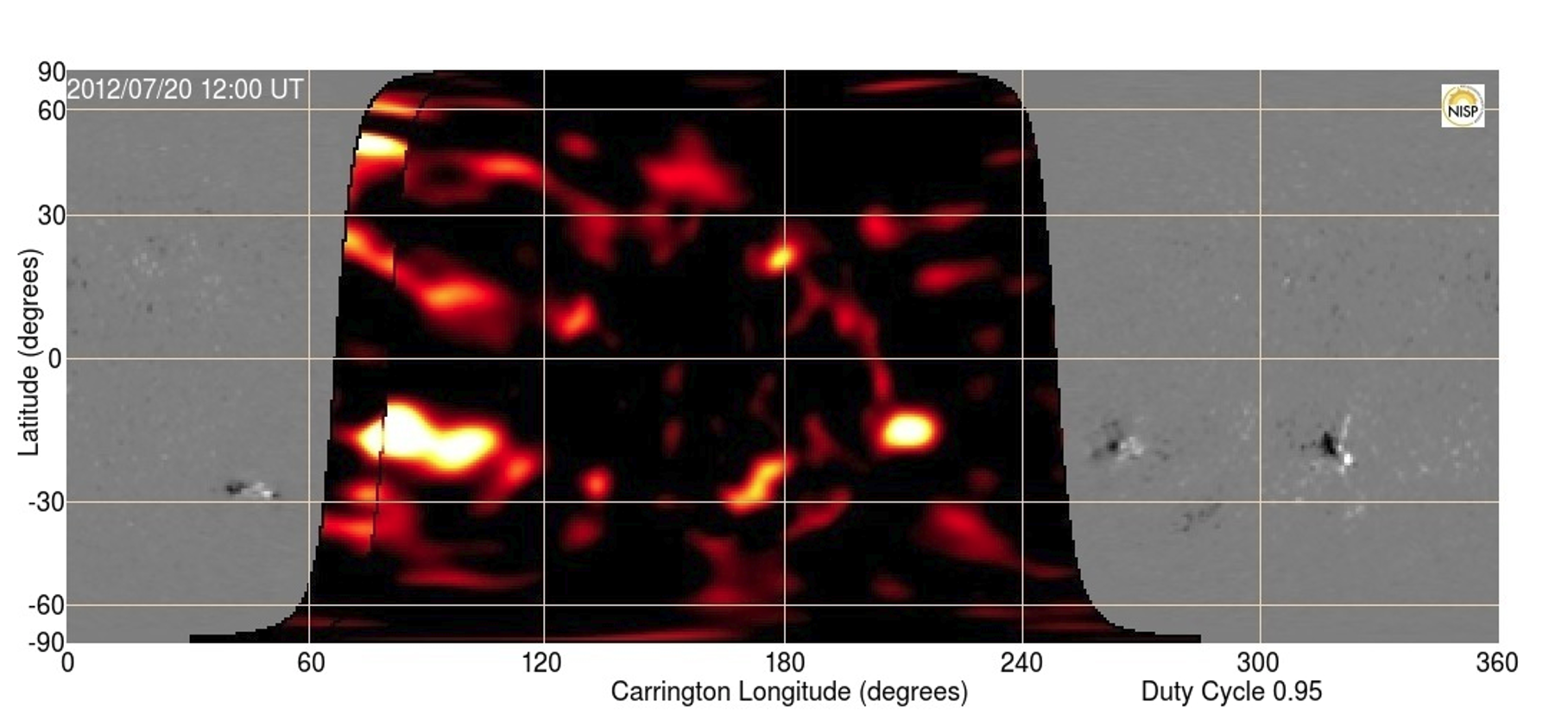}
                \label{cadence: 12hr}
            \end{subfigure}
            
            \vspace{1em} 
        
            \begin{subfigure}[b]{\textwidth}
                 \centering
                \includegraphics[width=0.95\textwidth]{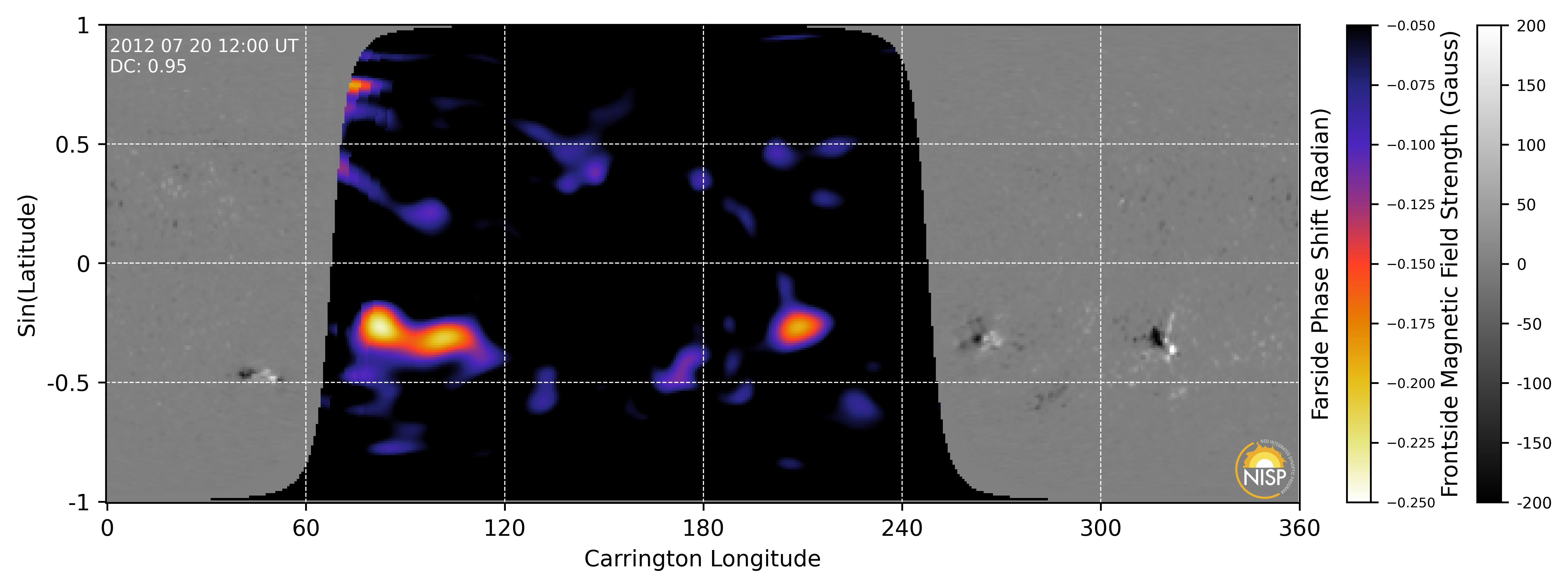}
                \label{cadence: 6hr}
            \end{subfigure}
            \caption{Comparison of legacy and new pipelines. The legacy pipeline ({\tt fqu}) map (a) consists of two individual 24 hour maps which are both 24 hours apart and are then averaged together, forming the 48 hour map. The new pipeline ({\tt f6u}) map (b) consists of five total 24 hour maps, each 6 hours apart. These maps are averaged together in order to provide the entire 48 hours of coverage.}
            \label{cadence comparison}
        \end{figure}

    \subsubsection[{Exclusion of Low Duty Cycle Maps}]{Exclusion of Low Duty Cycle Maps}
        \label{sec: Exclusion of low-duty cycle maps}
             
     As discussed in \citet{jainContinuousSolarObservations2021},  another potential source of noise is the  24 hour phase shift maps with low duty cycles. 
     Despite the high quality of the source data, low duty cycle maps often injected enough noise to compromise the usability of the final composites (Figure~\ref{cadence comparison}). To prevent contamination in the final composite maps, a new processing routine was developed and implemented. Maps with duty cycle under 80\% are systematically excluded as part of this process.
         



    \subsection[{Changes to Calibration Strategy: Variable Averaging}]{Changes to Calibration Strategy: Variable Averaging}
        \label{sec: Variable Averaging}

         When the 24 hour phase shift maps for each composite are combined in the legacy pipeline, each pixel is averaged by the total number of 24 hour phase shift maps within the composite. The result of this is a deadening signal at the most critical moment, when active regions are about to cross the Sun's east limb, before they return to face the earth.  This problem has been persistent across a variety of produced farside maps. This averaging equation is as follows: 
            
            \begin{center}
                $\bar{P} = \frac{1}{n_T} \sum_{i=1}^{n_T} P_i$
            \end{center}

        Where $\bar{P}$ is the average pixel value, ${n_T}$ is the total number of contributing maps, and ${P_i}$ is the pixel value from each of the contributing maps \\

        To prevent the signal dampening at the solar limb, the updated pipeline includes a variable averaging method, where each pixel is only averaged by the number of maps which contribute to that individual pixel. This is implemented as follows:

            \begin{center}
                $\bar{P} = \frac{1}{n_c} \sum_{i=1}^{n_c} P_c$
            \end{center}

        Where $\bar{P}$ is the average pixel value, ${n_c}$ is the number of maps contributing information to the signal, and ${P_c}$ is the pixel value from each of the contributing maps. This method has repeatedly proved successful in allowing for more accurate prediction of active region behavior during the critical period where farside active regions are crossing the limb to the earth side of the Sun. \\

    \subsection[{Magnetic Strength Map}]{Magnetic Strength Map}
        \label{sec: Magnetic Strength Map}
    
        \begin{figure} [h]
            \begin{center}
              \includegraphics[width=0.95\textwidth]{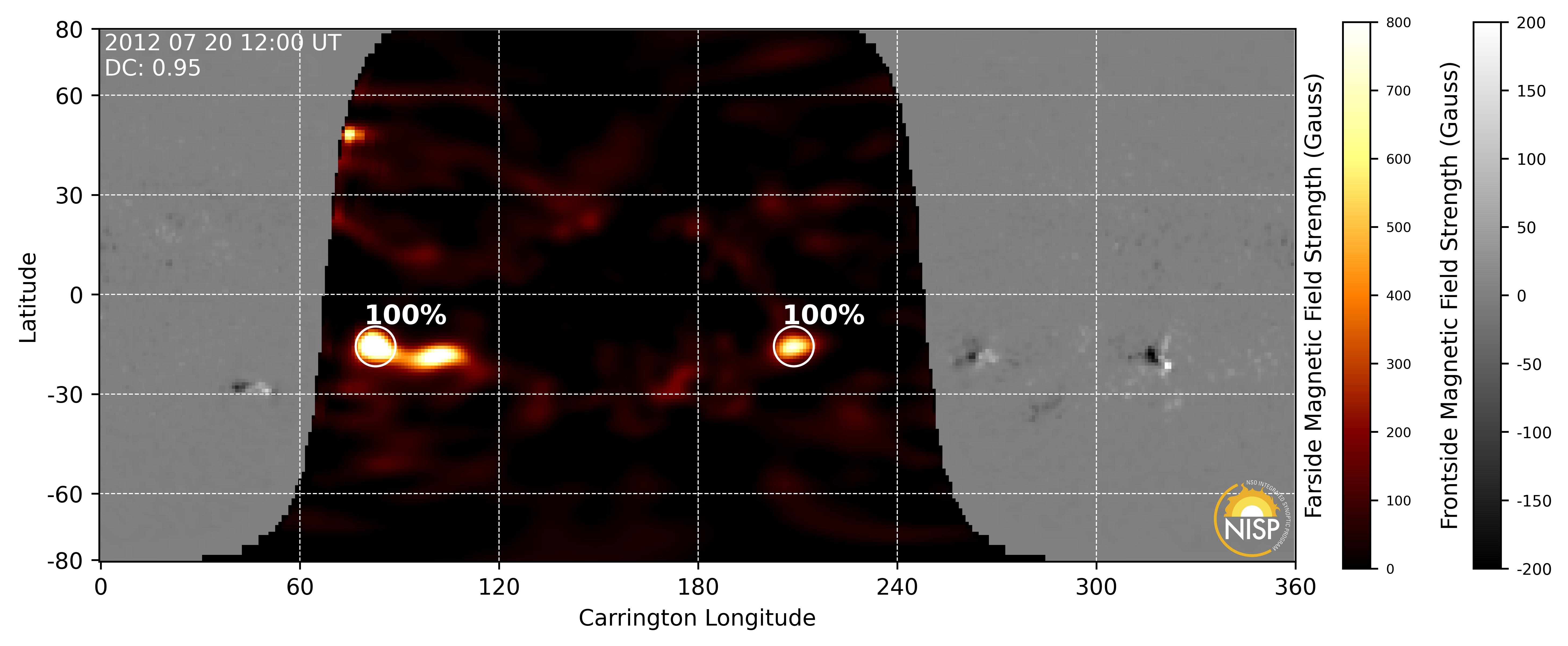}
                \caption{Magnetic Strength Map ({\tt mg} product) labeled with the active region probability for returning to the front-side.}
                \label{Mag Map}
            \end{center}
        \end{figure}
      
        One of the new data products from the farside pipeline is the production of magnetic strength maps, as shown in Figure~\ref{Mag Map}, derived from the phase shift maps. These maps are generated using a pixel-wise calculation of magnetic strength from the phase shift maps produced by the farside pipeline. The method of calculating magnetic strength from farside maps has also been updated. The original equation used to derive these magnetic strength maps is as follows:
        
            \begin{center}
                $$\text{$P_{mag}$} = \begin{cases} 
                    \sqrt{89.49} \times \left(\exp\left(\frac{\text{$P_{phase}$} \frac{\sqrt{2}}{\sqrt{6}}}{-0.014}\right) - 1\right) & \text{if } \text{val} <= 0 \\
                    0 & \text{if } \text{val} > 0
                \end{cases}$$
            \end{center}
        
         Where $P_{mag}$ is the pixel value for the magnetic map, and $P_{phase}$ is the phase shift value for the pixel at the corresponding [$ x , y$] position. This equation was updated due to further analysis of the data. As a consequence, the decision was made to update the magnetic strength calculation to the following equation using the same variables as the previous equation: 
    
        \begin{center}
            $$\text{$P_{mag}$} = \begin{cases} 
                \sqrt{566.44 \times \left(\exp\left(\frac{\text{$P_{phase}$}}{-0.0276}\right) - 1\right)} & \text{if } \text{val} <= 0 \\
                0 & \text{if } \text{val} > 0
            \end{cases}$$
        \end{center}

        These improvements have helped us to provide a considerably more accurate estimate of the magnetic strength of farside active regions.  
    
    \subsection[{Lat/Long Projection Maps}]{Lat/Long Projection Maps}
        \label{sec: Lat/Long Projection Maps}
    
        \begin{figure} [H]
            \begin{center}
              \includegraphics[width=0.95\textwidth]{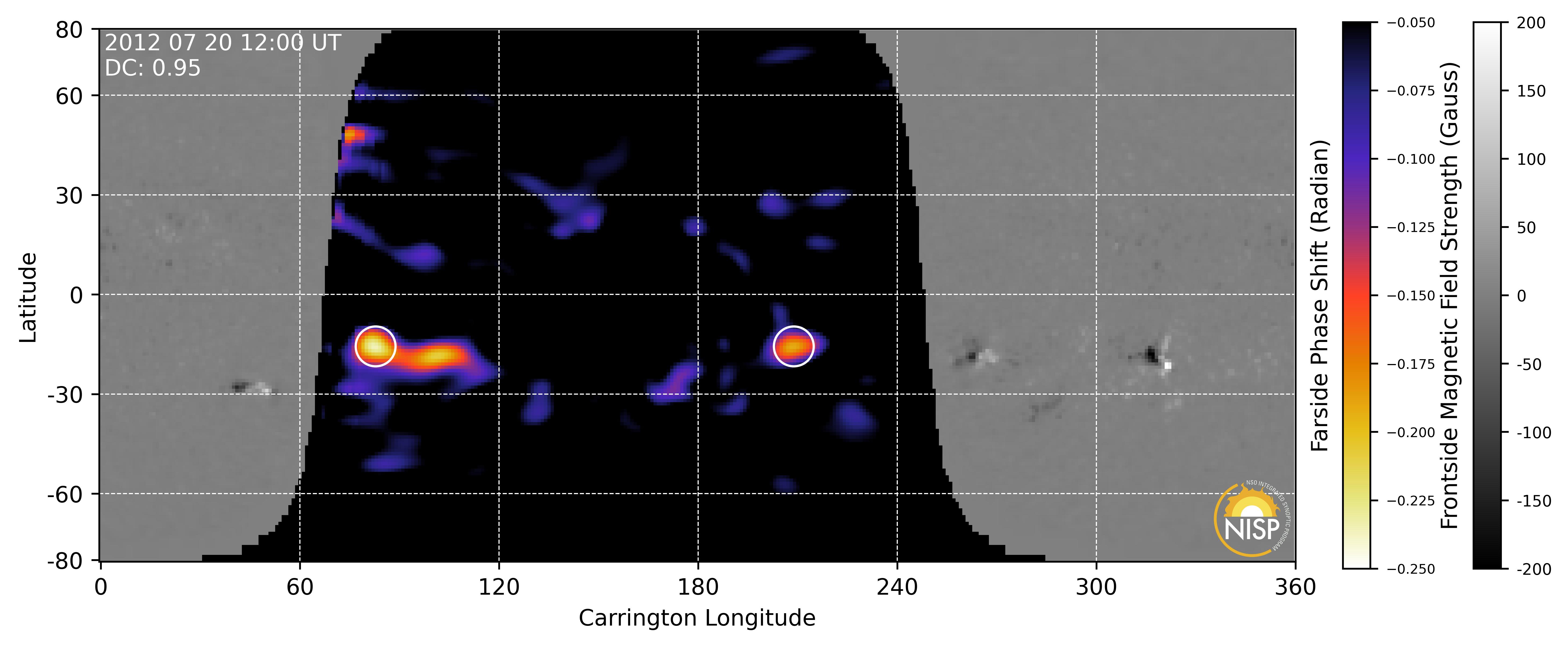}
                \caption{ Sample of farside phase shift maps in a Latitude/Longitude projection with a 1\degree\, resolution. These are stored as {\tt downsampled} products.}
                \label{DS Map}
            \end{center}
        \end{figure}
      
        Another new data products available in the updated pipeline are phase shift and magnetic strength maps that have been re-projected into [180 , 360] pixel Latitude/Longitude projections (Figure~\ref{DS Map}). The goal here is to provide a data product with a 1\degree\, resolution, resulting in a data product with enhanced usability for both the solar physics and space weather communities. Due to the stretching above the 80\degree\, latitudes, the visualizations produced are cut off at the upper limits of the map. \\

        As in the sine(latitude) projection maps, these composite maps are comprised at a 6 hour cadence with any 24 hour phase shift maps that fall under the 80\%\ duty cycle threshold being excluded from the final composite. Additionally, these maps are used to display active regions of interest for NSO publication to the wider space weather community. \\
        
        In this process, the Latitude/Longitude maps are first re-projected from a sine(Latitude)  to the linear latitude, deriving coordinates using the following equation, with \textit{y} being the y coordinate of the original pixel:
        
            \begin{center}
                $\text{Latitude}(\textit{y}) = \sin\left(\frac{\pi}{180} \times y\right)$
            \end{center}
    
        The resultant array is then interpolated from [200 , 500] to [180 , 360] using a linear interpolation strategy. Linear interpolation was chosen due to it having the best performance for our specific images when compared to multiple different methodologies. We compared the nearest neighbor, quintic, cubic, pchip, and slinear interpolation strategies. As tabulated in (Table \ref{Table: Interp. stats}), the slinear interpolation methodology had the lowest root mean squared error (RMSE), the highest peak signal-to-noise ratio (PSNR), and the second lowest multi-scale structural similarity index measure (MSSIM).
        
        \begin{table}[ht]
        \centering
            \caption{Comparison of RMSE, PSNR, and MSSIM across interpolation methods (Mean, Min, Max) for 401 maps.}
            \label{Table: Interp. stats}
            \begin{tabular}{l|ccc|ccc|ccc}
                \hline
                \textbf{Method} & \multicolumn{3}{c|}{\textbf{RMSE}} & \multicolumn{3}{c|}{\textbf{PSNR}} & \multicolumn{3}{c}{\textbf{MSSIM}} \\
                 & Mean & Min & Max & Mean & Min & Max & Mean & Min & Max \\
                \hline
                slinear & 0.004169 & 0.003316 & 0.009548 & 53.707 & 46.422 & 55.607 & 0.7998 & 0.7807 & 0.8206 \\
                pchip   & 0.004215 & 0.003347 & 0.009680 & 53.612 & 46.303 & 55.528 & 0.8013 & 0.7830 & 0.8221 \\
                cubic   & 0.004221 & 0.003349 & 0.009691 & 53.600 & 46.293 & 55.521 & 0.8544 & 0.8404 & 0.8673 \\
                quintic & 0.004227 & 0.003352 & 0.009708 & 53.587 & 46.278 & 55.514 & 0.8458 & 0.8328 & 0.8594 \\
                nearest & 0.004996 & 0.003994 & 0.011474 & 52.135 & 44.827 & 53.993 & 0.7946 & 0.7767 & 0.8163 \\
                \hline
            \end{tabular}
        \end{table}
        

    \subsection[{Global Projection}]{Global Projection}
        \begin{figure} [h]
            \begin{center}
              \includegraphics[width=0.95\textwidth]{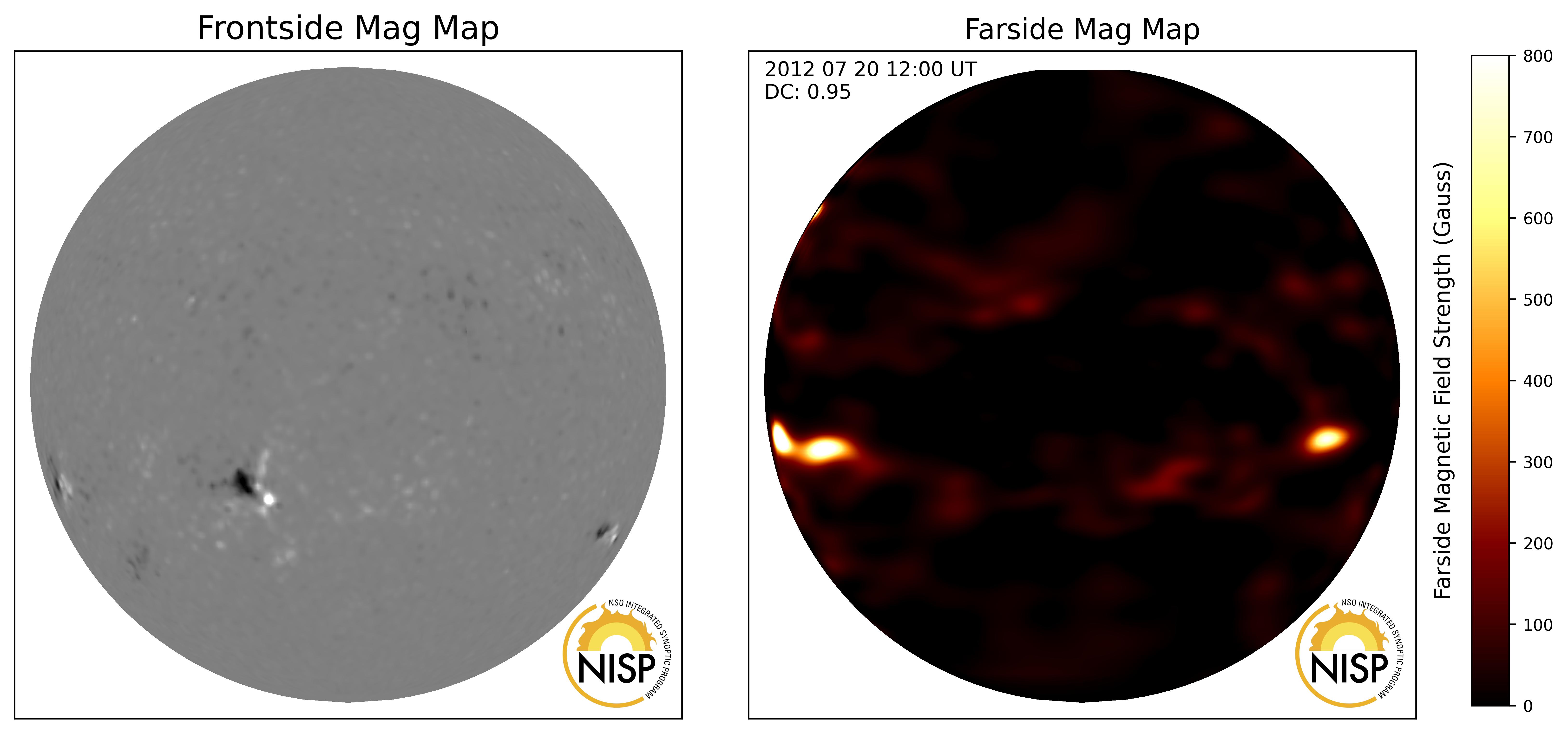}
                \caption{Spherical projection of farside magnetic strength maps, These are paired with a spherical projection for the composite frontside magnetograms ({\tt global} products) which are produced as part of the farside pipeline.}
                \label{Global Map}
            \end{center}
        \end{figure}

    Another new visual data product is a spherical and global projection of the farside phase shift and magnetic strength maps. These projections are derived from the initial [200 , 500] sine(Latitude) composite maps produced by the pipeline. The maps are projected using an algorithm provided by Gordon Petrie at the NSO. The observational angle and solar position are taken into effect to determine whether any given point is on the farside or Earth side. This series of trigonometric equations is given in  Appendix~A.1
    , where $\mathbf{\Lambda}$ is the latitude array and and $\mathbf{\Theta}$ is the co-latitude array. \\

     When combined with front-side magnetograms, these visualizations provide a powerful tool for tracking the persistence of solar active regions as they transit the solar disk. The goal is to develop a visualization product that is broadly accessible and capable of engaging audiences beyond the scientific community, thereby enhancing understanding of the farside project’s utility. 

     Further to this end, fits files of these fully calibrated line of site projections have been produced in a 540x540 resolution for fully calibrated magnetic field strength maps and fully calibrated phase shift maps

    \section[{ Code Structure Outline}]{Code  Structure Outline}
        \label{sec: Code Structure}

        In order to accommodate the necessary changes to the updated farside pipeline without fundamentally disturbing the performance of existing data production, a new end phase to the farside pipeline was developed. This pipeline is known as the '{\tt f6}' pipeline after the 6 hourly cadence of the 24 hour phase shift maps used in the final maps. This pipeline uses the same source data as the '{\tt fq}' legacy pipeline, but uses a higher temporal cadence of 24 hour phase shift maps in addition to a different calibration process to produce the final composite maps seen by the end users. This section provides a general outline of the differences between the legacy '{\tt fq}' pipeline, and the updates '{\tt f6}' pipeline. The specifics of the code structure and build processes can be found in the wiki.

        \subsection{FQI Filter and Image Merge}
            \label{sec: Image Merge}

            The filter for anomalous FQI images examines images just before they are send to the oQR repository. This filter is a daemon which runs at a cadence of one minute to check if any new fqi images have been submitted to the input repository. If new files have been created, the filter will run a check on the new images before they are sent to production. All images that are scanned are given the following keywords in the header:
            
            \begin{verbatim}
                ML_VRSN  - Machine Learning (ML) algorithm version number 
                REC_QUAL - ML quality flag (0 means good)                 
                PROBGOOD - ML probability of good data                    
                PROBANOM - ML probability of anomalous data
            \end{verbatim}
            
            These headers provide the information necessary for both our pipelines, and the data processing routines of any interested parties, to use the results of our algorithm to produce cleanly merged quick-reduce dopplergrams. If an updated algorithm is applied to re-label images, it will be noted in the ML\_VRSN keyword.

            The details of the machine learning process can be found in \citep{creelmanAnomalyDetectionGONG2024} with the efficacy of implementation being noted in \citep{Jain2025SoPh}.
    
        \subsection[{Increased Cadence 24 hour phase shift maps }]{Increased Cadence 24 hour phase shift maps}
            \label{sec: Cadence Increase}

        The existing legacy farside pipeline produced a phase map (fqm product) every 12 hours. To facilitate more averaging, in the new pipeline a phase map is produced every 6 hours. To do this, the new system is run on the same input data (merged fqi Dopplergrams) as the existing system, but more frequently. No changes were made to the fundamental algorithm that reads merged Dopplergrams and writes phase maps (although such changes are being discussed). Fundamentally, the only change was to the making of the lists of input files, which now have to be generated more frequently.

        \subsection[{Calibrated Maps: New Mapping Structure and Processes}]{Calibrated Maps: New Mapping Structure and Processes}
            \label{sec: New Mapping Structure and Processes}
                    \begin{figure} [h]
                        \begin{center}
                          \includegraphics[width=0.9\textwidth]{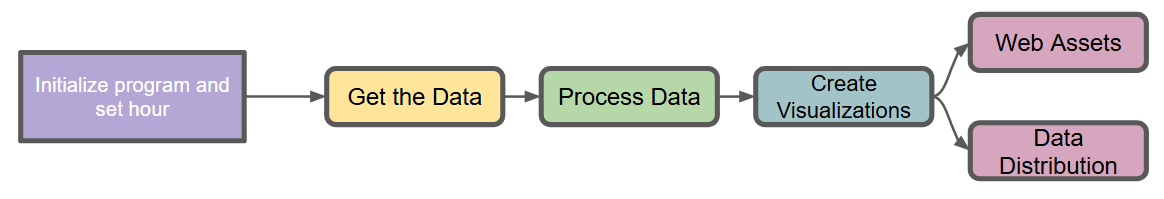}
                            \caption{ Diagram of the code structure for the end of the {\tt f6u} pipeline.}
                            \label{f6u_structure}
                        \end{center}
                    \end{figure}

                The major final component of the {\tt f6u} pipeline is the updated mapping routine. Overall, the components of the pipeline ending are called in sequence by a shell script. This script calls the items detailed in figure \ref{f6u_structure} at a 12 hour cadence. The script determines the exact time to produces, finds the applicable maps to integrate into the composite products, and then creates a composite map. Only maps generated within a 24 hour window with a duty cycle greater than 0.8 will be included in the final composite. All other maps will be excluded in order to minimize noise in the final farside mapping product. \\

                Once a composite map is formed, the find\_candidate script is run to identify farside active regions of particular interest, generating a list of their characteristics and coordinates. Included All active regions with a probability over 95\% are automatically included in the labeled products. Products lower than this, however, must meet two criteria: a frontside return probability score of at least 70\%  and must be detected for at least two consecutive maps. The resulting data is compiled in the {\texttt *\_pub.csv} file, which contains the following information:

                \begin{verbatim}
        C. Long             - Active region center of mass Carrington longitude
        Latitude            - Active region center of mass Latitude
        Phase-Min           - The minimum phase strength of the active region
        Phase-Strength      - The total phase strength of the active region
        Prob.               - Likeliehood that the active region returns to face earth
        Effective Area      - Area of the active region
        Magnetic Flux       - Total sagnetic strength of the active region
        Magnetic Max        - Maximum magnetic strength of the active region
        Date of Return      - Date when the active region center of mass will cross the solar limb
        Front-Side-Number   - NOAA Active Region number based on matching coordinates
                \end{verbatim}

                A copy of the csv, along with a list of After all potential Active regions have been identified, new map products are produced. These products include farside and phase shift maps produced in sin(longitude) and Carrington longitude projections, as well as global projection forms. Additionally, labeled maps for phase shift and magnetic strength maps are produced in Carrington  longitude projection, with any relevant active regions identified with a circle. These maps are available to the general public via the NISP data distribution system.

\section[{Results}]{Results}
    \label{sec: Results}
    
   In this section, we present results demonstrating improvements across two key areas. The first area of improvement is the enhanced near-limb detection of active regions transitioning from the Sun's far side to the Earth-facing side, a critical phase where previous calibration methods proved insufficient. The updates detailed in this report have yielded a significantly more robust product, enabling more effective tracking of solar active regions during this transitional period.\\

    The second set of results highlights image quality metric improvements across 7,757 fqu and f6u farside maps. This analysis compares the variance, spatial frequency, and sharpness ratio for fqu and f6u maps of the same date. These measures help us to quantitatively determine the efficacy of the updated calibration techniques implemented in this iteration of the Farside pipeline.

    \subsection[{Near-Limb Detection Improvements}]{Near-Limb Detection Improvements}
        \label{sec: Near-Limb Detection Improvements}

        \begin{figure}[htbp]
            \begin{center}
            \begin{subfigure}[t]{0.4\textwidth}
                \centering
                \includegraphics[width=\linewidth]{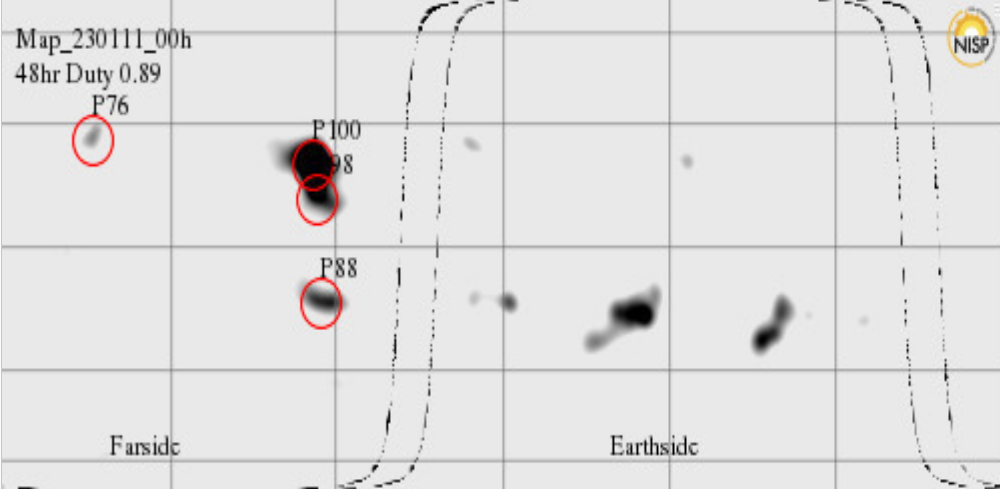} 
                \caption{mrfqu230111t0000}
                \label{subfig: fqu1}
            \end{subfigure}
            \hfill
            \begin{subfigure}[t]{0.4\textwidth}
                \centering
                \includegraphics[width=\linewidth]{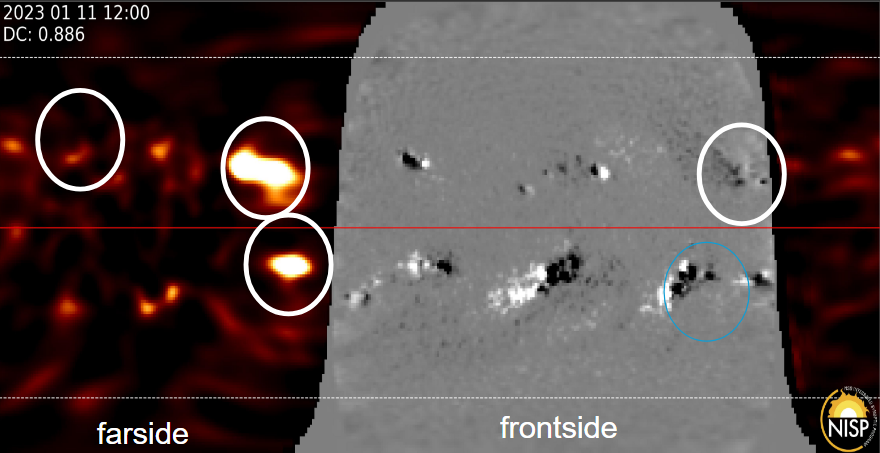} 
                \caption{mrf6u230111t0000}
                \label{subfig: f6u1}
                
            \end{subfigure}
        
            \vspace{0.5em} 
        
            \begin{subfigure}[t]{0.4\textwidth}
                \centering
                \includegraphics[width=\linewidth]{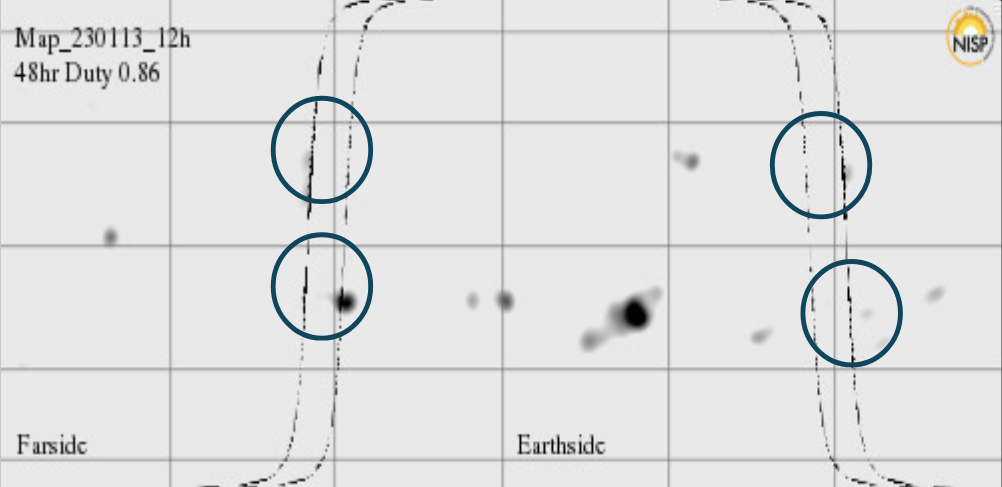} 
                \caption{mrfqu230113t1200}
                \label{subfig: fqu2}
            \end{subfigure}
            \hfill
            \begin{subfigure}[t]{0.4\textwidth}
                \centering
                \includegraphics[width=\linewidth]{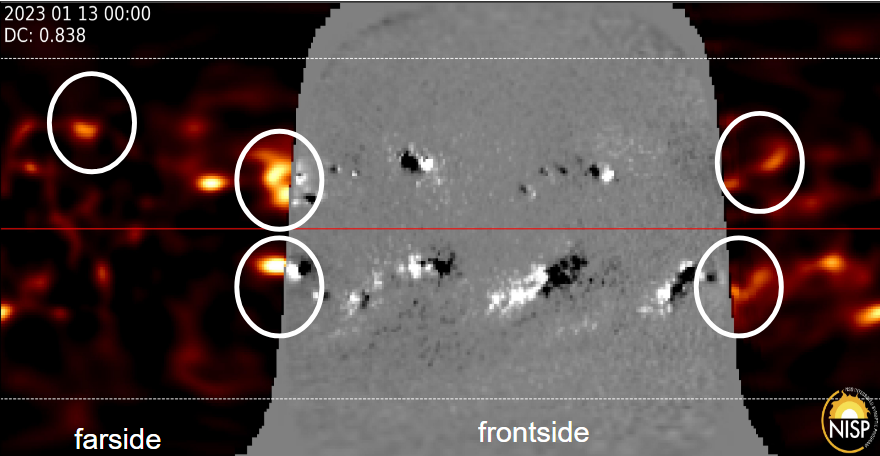} 
                \caption{mrf6u230113t1200}
                \label{subfig: f6u2}
            \end{subfigure}
            
            \caption{Figure showing the improvements in edge detection between the legacy pipeline (a/c) and the updated pipeline (b/d) in Farside magnetic strength maps. Notice the persistence of the active regions in the updated maps, while the identical active regions disappear rite before emergence in the old products.}
            \label{fig: Edge Detection Sample}
            \end{center}
        \end{figure}

        One of the most critical shortcoming of the legacy pipeline is the reduction in seismic signal when active regions are nearing the East limb. Given that the main goal of farside mapping is to provide accurate tracking of the behavior and emergence of farside active regions, losing sight of these regions just as they are about to return to the Earth side of the Sun is operationally unsatisfactory. The new calibration methodologies in the updated pipeline have provided significant improvements in the visibility of farside active regions during this critical period, as shown in Figure~\ref{fig: Edge Detection Sample}. \\

        While numerous examples could demonstrate this improvement, the active regions marked in Figure~\ref{fig: Edge Detection Sample} are representative of the overall
        improvements. Here we see that 
        he enhanced averaging methodologies have substantially mitigated the signal attenuation effect near the solar limbs. We notice that the active regions  marked in Figure~\ref{subfig: fqu1} completely disappear when they reach the East limb in Figure~\ref{subfig: fqu2}. This is directly contrasted with the same active regions being identified in Figure~\ref{subfig: f6u1} persisting through and still being visible while part of them have already passed to the frontside in Figure~\ref{subfig: f6u2}. This is simply one illustration of the efficacy of the newly implemented calibration measures.

    \subsection[{Improvements in Image Quality}]{Improvements in Image Quality}
        \label{sec: Improvements in Image Quality}

        While near-limb detection is clearly a critical improvement for the updated pipeline, the question remains as to how the overall quality of the farside maps has been improved with the updated calibration methodology. To this end, we have analyzed variance, sharpness ratio, and Shanon entropy for 7,757 temporally identical farside maps. These metrics are meant to show the progress towards the overall goal of reducing background noise in end product composite maps. By looking at the changes in these 3 metrics for the legacy and new metrics, we can clearly see that the implemented measures have had a significant and quantitative impact on the quality of the farside maps produced by the NSOs GONG program 

        \subsubsection[{Image Variance}]{Image Variance}
            \label{sec: Image Variance}

            Variance is the measure of central distribution around the mean values of an image. In farside phase shift maps, noisy maps see a wider range of values than quiet maps and are thus show a larger variance for the map. Variance of an image is calculated with the following equation:
            
                \begin{equation}
                    \text{Variance} = \frac{1}{N} \sum_{i=1}^{N} (x_i - \mu)^2
                \end{equation}
            
            Where $N$ is the total number of pixels in the image, $x_{i}$ is the pixel value at that specific location, and $\mu$ represents the mean pixel intensity for the image \citep{aparnaImageVariance2017}. As previously mentioned, we calculated variance for 7,757 temporally identical farside maps in order to compare the overall impact of the updated calibration sequence implemented in the new farside pipeline. \\

\begin{figure}[t] 
                    \label{fig: Variance Performance}       
                    \centering
                    \begin{subfigure}{0.75\linewidth}
                        \centering
                        \includegraphics[width=\linewidth]{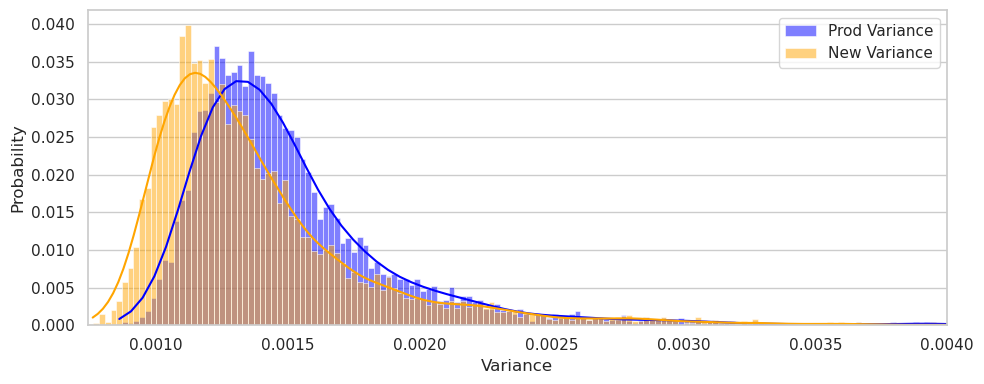} 
                        \label{fig: variance histogram}
                        \caption{Histogram of the variance distributions for the legacy and updated farside calibration methodologies. }
                    \end{subfigure} 
                    \begin{subfigure}{0.65\linewidth}
                        \includegraphics[width=\linewidth]{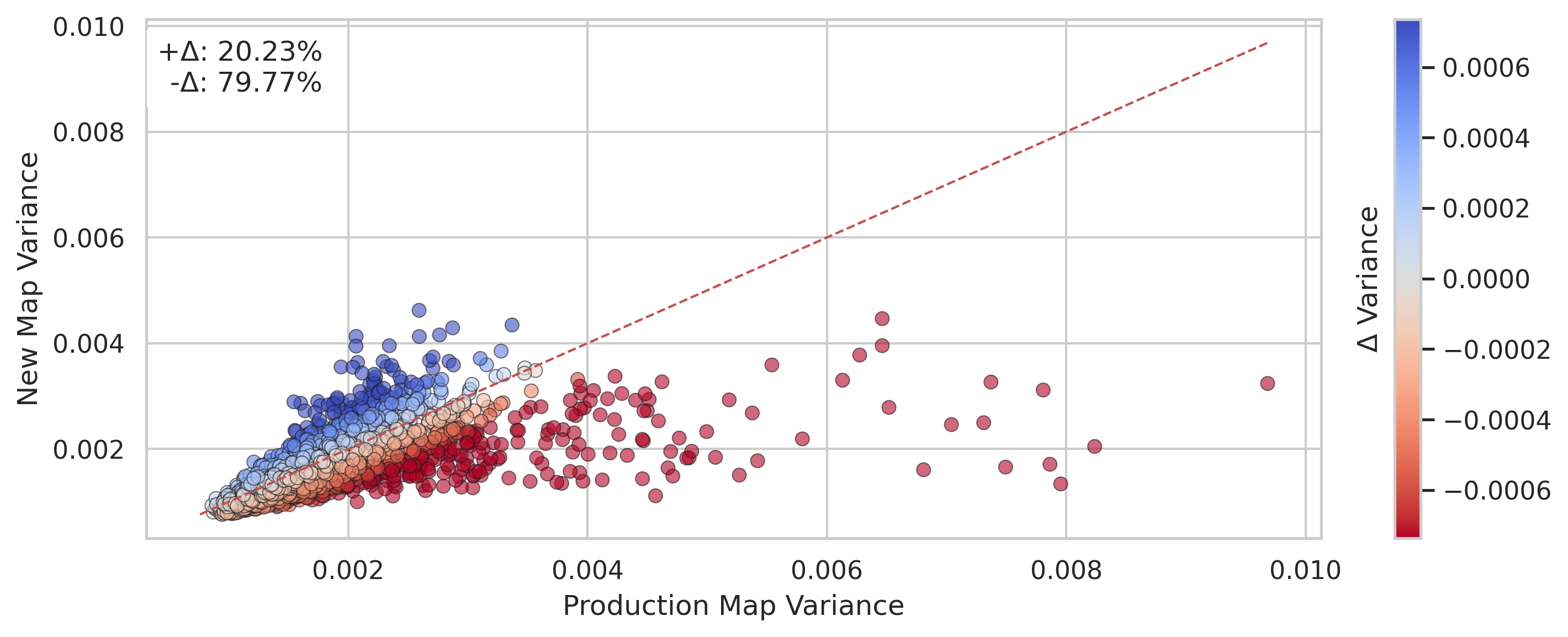} 
                        \label{fig: variance_scatter}
                        \caption{Scatter plot showing the variance of temporally identical maps from the legacy and updated pipeline. The coloration is taken from the magnitude of the difference between the two pipelines }
                    \end{subfigure}                
                    \caption{Variance Statistics results for 7,757 temporally identical maps produced by the legacy and updated pipelines}\label{fig: variance performance}
                \end{figure}

            The comparison of old and new maps shows a marked and significant increase in image variance, as seen in Figure~\ref{fig: variance performance}(a). Figure~\ref{fig: variance performance}(b) shows a marked reduction in the overall distribution of variance for the examined maps. The mean variance saw a reduction from 0.00153 to 0.00139. with 79.8\% of maps seeing a reduction in variance. Additional the media variance dropped from 0.00141 to 0.00126. Additionally, we can see that no maps exceed a variance of 0.0046. This variance ceiling is significant as it shows a dramatic reduction in noisy and unusable maps being produced by the updated pipeline. Overall, the updated calibration processes have resulted in a far more reliable farside mapping products than previous iterations of the pipeline.    
                       
        \subsubsection[{Shannon Entropy}]{Shannon Entropy}
            \label{sec: Shannon Entropy}

            The Shannon Entropy of an image a metric used to describe the amount of uncertainty and information within a particular image \citep{aMathematicalTheoryOfCommunication}. A higher entropy indicates more noise within a given image. The indirect observational nature of the farside product makes any reduction in entropy within the maps themselves extremely valuable to their overall utility to the space weather community.  Shannon entropy is calculated with the following equation:

                \begin{equation}
                    H(X) = -\sum_{i=1}^{n} P(x_i)\log_2P(x_i)
                \end{equation}
                

            Where $H(X)$ denotes the entropy of a discrete random variable, and $P(x_i)$ is the probability of x within the image. 
            This calculation was run on the sample of 7,757 temporally identical maps produced by both methodologies for the following results:\\
        
                \begin{figure}[t] 
                    \centering
        
                    \begin{subfigure}[t]{0.65\linewidth}
                        \centering
                        \includegraphics[width=\linewidth]{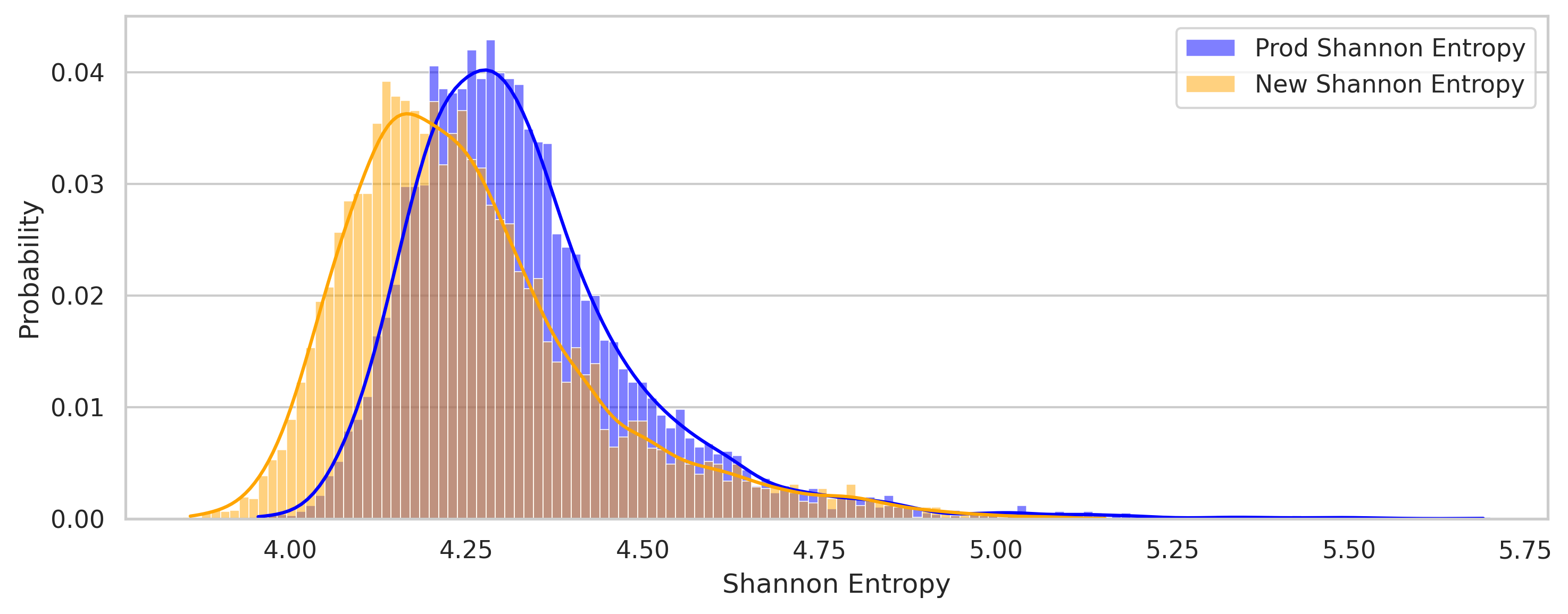} 
                        \caption{Histogram showing distribution of Shannon Entropy values for legacy and updated pipeline maps.}
                        \label{fig: SE Hist}
                    \end{subfigure}
                    
                
                    \begin{subfigure}[t]{0.65\linewidth}
                        \centering
                        \includegraphics[width=\linewidth]{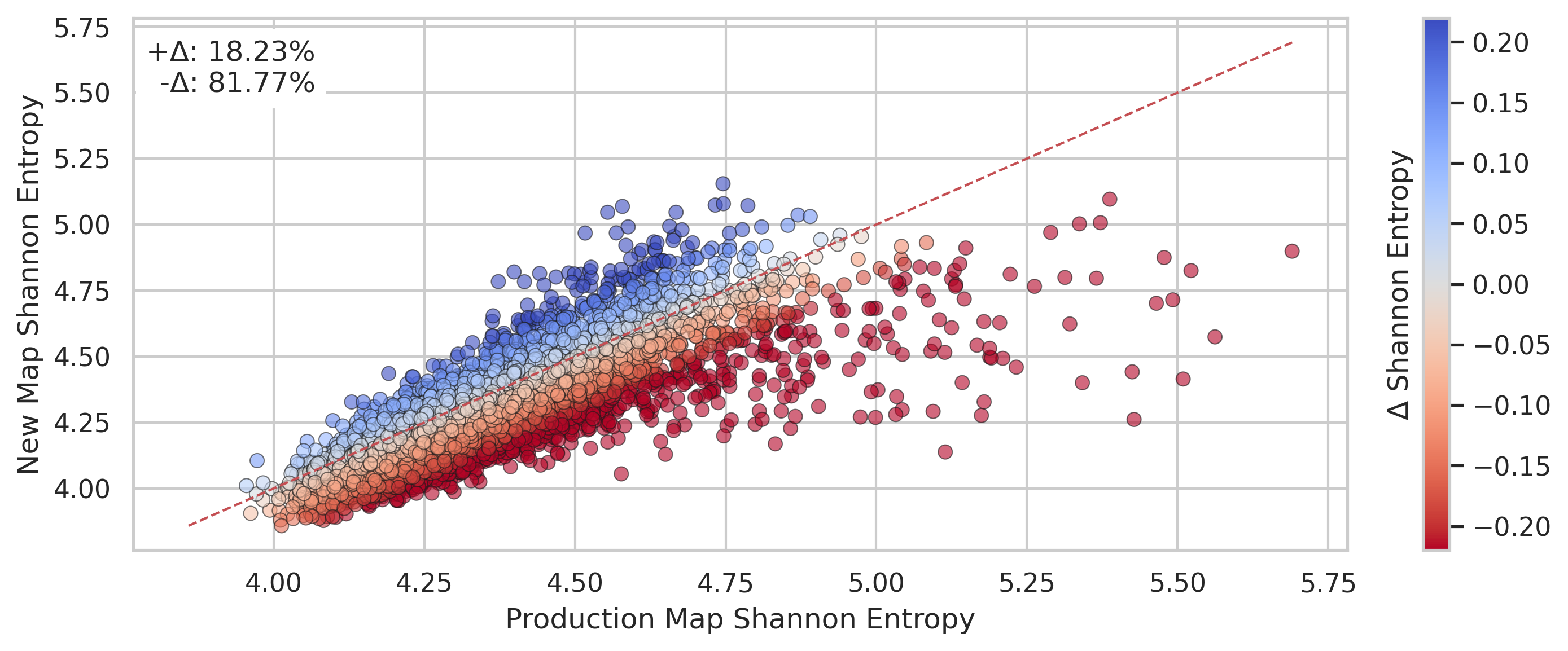} 
                        \caption{Scatterplot comparing the Shannon Entropy values for temporally identical maps generated by the legacy and updated pipelines.}
                        \label{fig: SE Scatter}
                    
                    \end{subfigure}
                    \caption{Shannon Entropy Values for 7,757 temporally identical maps produced by the legacy and updated pipelines}
                    \label{fig: Shannon Entropy Performance}
                \end{figure}

            Figure \ref{fig: Shannon Entropy Performance} shows a significant entropy reduction in maps produced by the two pipelines, with 81.8\% of all maps seeing a reduction in entropy due to the newly applied methodologies. This includes a mean reduction from 4.33 to 4.26, with the median entropy falling from 4.30 to 4.22. This dramatic reduction in image entropy in the fully calibrated maps quantitatively shows us that the new calibration methodologies are resulting in significantly less noise in the final end-product farside maps.


        \subsubsection[{Spatial Frequency}]{Spatial Frequency}
            \label{sec: Spatial Frequency}
            
            The spatial frequency of an image measures the intensity difference between adjacent pixels (\citep{lahiriOpticalCoherence2016}). This gives us a measure of consistency within an image, with a higher score indicating more frequent and more severe outliers within a given image. In our maps, a higher level of spatial frequency would indicate a noisier product, thus a lower frequency would indicate and overall higher quality map product. This metric is calculated with the following equation:

                \begin{equation}
                    \text{Spatial Frequency} = \sqrt{\text{mean}(\text{diff\_row}^2) + \text{mean}(\text{diff\_col}^2)}
                \end{equation}

                \begin{figure}[h] 
                    \centering
        
                    \begin{subfigure}[t]{0.65\linewidth}
                        \centering
                        \includegraphics[width=\linewidth]{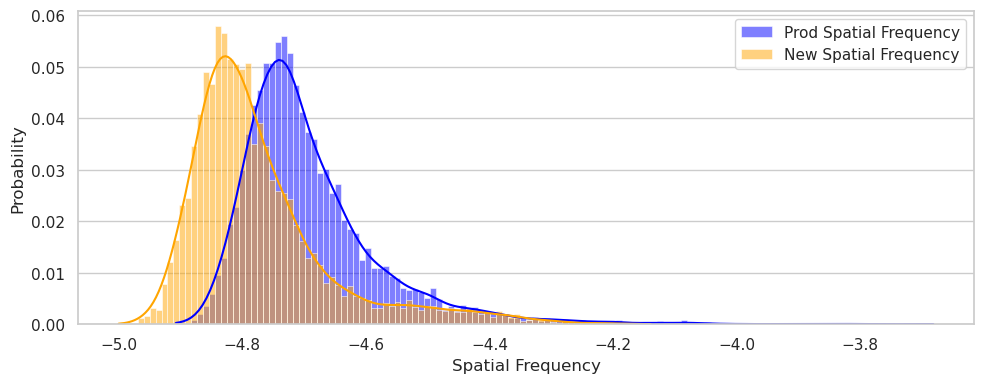} 
                        \caption{Histogram showing distribution of Spatial Frequency for legacy and updated pipeline maps.}
                        \label{fig: SF Hist}
                    \end{subfigure}                  
                    \begin{subfigure}[t]{0.65\linewidth}
                        \centering
                        \includegraphics[width=\linewidth]{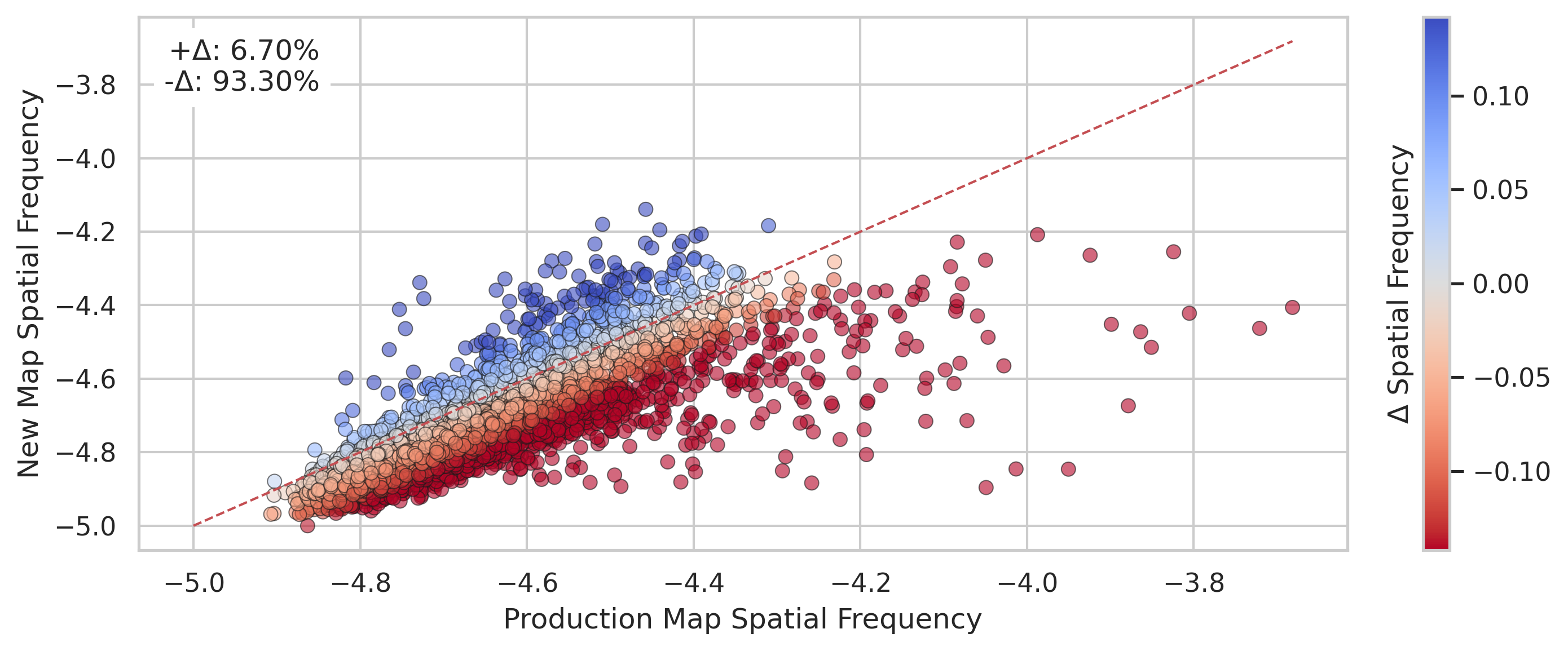} 
                        \caption{Scatterplot comparing the Spatial Frequency for temporally identical maps generated by the legacy and updated pipelines.}
                        \label{fig: SF Scatter}                
                    \end{subfigure}
                    \caption{Spatial Frequency Ratio results for 7,757 temporally identical maps produced by the legacy and updated pipelines }
                    \label{fig: Spatial Frequency Performance}
                \end{figure}

            Figure \ref{fig: Spatial Frequency Performance} shows the dramatic decrease in spatial frequency, with 93.3\% of the total maps analyzed seeing a reduction. This includes a mean reduction from -4.77 to -4.69, with the median value reducing from -4.80 to -4.72. This reduction shows more consistency within the produced maps, again showing that the newly implemented calibration methodologies have a marked and significant impact on the overall quality of the maps being reduced. 

\section[{Discussion}]{Discussion}
    \label{sec: Discussion}

    It is clear that the improved calibration measures, including the increased temporal cadence detailed in this report have resulted in a far better products. A list of new products is given in Appendix A.2. These improved farside maps generated by the GONG network deliver enhanced reliability and precision in phase shift mapping, benefiting space weather studies and forecasts, allowing for better near-real time monitoring of active regions during the periods when active regions pass to the far side of the Sun where they cannot be continuously directly observed.The capacity to implement enhancements like these represents a key advantage of ground-based solar observatory networks such as GONG. Notably these updates are part of a broader set of planned improvements to the farside pipeline. We anticipate that integrating machine learning algorithms into the map production process will further increase the reliability and accuracy of the resulting products for end-users.

    \subsection[{Future Work}]{Future Work}
        \label{sec: Further Improvements}
        While the measures and methods outlined in this technical report represent a significant improvement to the products produced by the NSO's farside pipeline, there are still further improvements that we would like to see implemented to the project. These improvements would further improve the clarity of the produced maps, provide more accurate identification of farside active regions, and would result in new data products for the larger solar science community.

    \subsubsection[{Find Candidate}]{Find Candidate}
        \label{sec: Find Candidate}
    
        One area with particular opportunity for improvement is the find\_candidate regime that is used in the legacy pipeline. As outlined previously, the {\tt find\_candidate} function is used to identify solar active regions on the far side of the Sun, pair them with their associated NOAA numbers, and calculate their size, and strength, before assigning them a date of return. While this traditional stochastic methodology provides adequate performance for larger active regions, it often misses smaller and emerging areas of interest. The ability to accurately and precisely identify such regions would be an immeasurable step forward for the field of space weather forecasting, allowing more advanced notice and tracking of solar active regions before they face the Earth.\\

        To this end, we are developing a machine learning algorithm designed to enhance the precision of active region identification and tracking. The chosen network architecture is a long short-term memory (LSTM) model, informed by prior work on tracking farside active regions \citep{broockFarNetIIImprovedSolar2022,felipeImprovedDetectionFarside2019}. While these methods alone are promising, we have also developed a ground-truth dataset of farside active regions by comparing maps from the new {\tt 6u} pipeline to maps collected by the STEREO mission \citep{hamadaFarsideActiveRegions2024}. Our goal is to create a neural network specifically tailored the NSO {\tt f6u} pipeline to identify and track farside active regions with far more precision and accuracy than the current methodologies provide. We have made progress in this area  \citep{Hamada2025a}, and the algorithm is currently undergoing validation.
    
    \subsubsection[{Farside Signed Magnetic Flux Maps from Phase Shifts}]{Farside Signed Magnetic Flux Maps from Phase Shifts}   
        Another goal of the farside research group is to develop a methodology for generating farside signed magnetic flux maps from the current phase shift products. Our plan is to train an image translation \citep{isolaImagetoImageTranslationConditional2018} and a super-resolution \citep{shiRealTimeSingleImage2016,ledigPhotoRealisticSingleImage2017} network on the ground-truth phase shift maps by \citet{hamadaFarsideActiveRegions2024}. The goal of this pipeline would be to provide the scientific community with magnetic maps for the Sun's farside without the need for expensive, inconsistent, and logistically challenging direct observation missions \citep{kaiserSTEREOMissionOverview2008,mullerSolarOrbiterMission2020}.                                                                                                                                                          
     \subsubsection[{Dopplergram Calculation Temporal Window Update}]{Dopplergram Calculation Temporal Window Update}
        \label{sec: length}     
       The current system runs on a 24 hour sequence of Dopplergrams to represent a 24 hour period. Since the travel time for waves from the farside to the frontside of the Sun is about 3.5 hours, it is arguable that the 24 hour temporal window should be extended 3.5 hours both earlier and later for a 31 hour temporal window with the same central time. In addition, It is worth testing with the time series of 48 hours. 
       
        \subsubsection[{Green's Function Revisions}]{Green's Function Revisions}
        \label{sec: skips}
       As stated, the current system repeats Step 4 (Section~\ref{sec: Legacy Composite Map Creation (fsWeb)}) three times to encompass three skip patterns. It is arguable that the analysis should not be skip pattern dependent, rather, one analysis that encompasses all skip patterns should be done. This is an involved problem that requires solar modeling to estimate the Green's function (essentially, the impulse response of the Sun) independent of skip pattern.\\
          

     \addcontentsline{toc}{section}{\bf Acknowledgments}
    {\bf Acknowledgments:} We thank Gordon Petrie for providing the algorithm used in global projection. This work utilizes GONG data obtained by the NSO Integrated Synoptic Program, managed by the National Solar Observatory, which is operated by the Association of Universities for Research in Astronomy (AURA), Inc. under a cooperative agreement with the National Science Foundation (NSF) and with contribution from the National Oceanic and Atmospheric Administration (NOAA). The GONG network of instruments is hosted by the Big Bear Solar Observatory, High Altitude Observatory, Learmonth Solar Observatory, Udaipur Solar Observatory, Instituto de Astrof\'{\i}sica de Canarias, and Cerro Tololo Interamerican Observatory. This work was supported, in part by the  Windows-on-the-Universe Multi-Messanger Astrophysics (WoU MMA) project via  amendment to the NSF Cooperative Agreement, Supplemental Funding Request (SFR) Amendment \#040.
   \addcontentsline{toc}{section}{\bf References} 
\bibliography{references}   

\pagebreak
\addcontentsline{toc}{section}{\bf Appendix}
\begin{center}
{\bf \Large Appendix}
\end{center}

    \addcontentsline{toc}{subsection}{\bf A.1: Global Projection Calculation}{\bf \large A.1: Global Projection Calculation}
        \label{sec: Global Projection Calculation}

Algorithm 
used to compute the global projection of the calibrated maps. The arrays $\mathbf{\Lambda}$ and $\mathbf{\Theta} $ represent latitude and co-latitude, respectively. All other symbols have their conventional definitions.

        \begin{center}
            $\alpha = \frac{\pi}{180}\text{$L_0$}   \hspace{2cm} \angle_B = \frac{\pi}{180}\text{$B_0$}$
            
            \vspace{0.5cm}
            $\mathbf{\Lambda} = \arcsin\left(\frac{\sin\left(\mathbf{\Lambda}\right)} {\frac{\pi}{180}} \right)$

            \vspace{0.15cm}
            $\mathbf{\Theta} = 90 - \mathbf{\Lambda}$

         \vspace{0.5cm}
        \[ 
            \mathbf{\Lambda} = \begin{bmatrix}
            \lambda_1 & \lambda_2 & \dots & \lambda_n \\
            \lambda_1 & \lambda_2 & \dots & \lambda_n \\
            \vdots & \vdots & \ddots & \vdots \\
            \lambda_1 & \lambda_2 & \dots & \lambda_n \\
            \end{bmatrix}
        \]
        
         \vspace{0.15cm}   
            $\mathbf{\Theta} = \begin{bmatrix}
            \theta_1 & \theta_1 & \dots & \theta_1 \\
            \theta_2 & \theta_2 & \dots & \theta_2 \\
            \vdots & \vdots & \ddots & \vdots \\
            \theta_m & \theta_m & \dots & \theta_m \\
            \end{bmatrix}$

        \vspace{0.5cm}
        $\Lambda_{i,j} = \lambda_j \quad \text{for } i = 1, \dots, m \text{ and } j = 1, \dots, n$
        
        \vspace{0.15cm}
        $\Theta_{i,j} = \theta_i \quad \text{for } i = 1, \dots, m \text{ and } j = 1, \dots, n$

        \vspace{0.5cm}
        $ \cos_{\Theta_{i,j}} = \cos\left(\Theta_{i,j} \times \frac{\pi}{180} \right)$

        \vspace{0.15cm}
        $ \sin_{\Theta_{i,j}} = \sin\left(\Theta_{i,j} \times \frac{\pi}{180} \right) $

        \vspace{0.15cm}
        $\zeta_{\Theta_{i,j}} = \cos_{\Theta_{i,j}} $

        \vspace{0.5cm}
        $ \cos_{\Lambda_{i,j}} = \cos\left(\Lambda_{i,j} \times \frac{\pi}{180} \right)$

        \vspace{0.15cm}
        $ \sin_{\Lambda_{i,j}} = \sin\left(\Lambda_{i,j} \times \frac{\pi}{180} \right) $

        \vspace{0.5cm}
        $X_{i,j} = \cos_{\Lambda_{i,j}} \times \sin_{\Theta_{i,j}} $

        \vspace{0.15cm}
        $Y_{i,j} = \sin_{\Lambda_{i,j}} \times \cos_{\Theta_{i,j}}$

        \vspace {0.5cm}
        $Xr_{i,j} = X_{i,j} \times \cos(\alpha) + Y_{i,j} \times \sin(\alpha) $

        \vspace{0.15cm}
        $ Yr_{i,j} = Y_{i,j} \times \cos(\alpha) - Xr_{i,j} \times \sin(\alpha)$

        \vspace{0.15cm}
        $\zeta r_{i,j} = \zeta_{i,j}$

        \vspace{0.5cm}
        $Xt_{i,j} = Xr_{i,j} \times \cos(\angle_B) + \zeta r_{i,j} \times \sin(\angle_B)$

        \vspace{0.15cm}
        $Yt_{i,j} = Yr_{i,j} $

    `   \vspace{0.15cm}
        $\zeta t_{i,j} = \zeta r_{i,j} \times \cos(\angle_B) - Xr_{i,j} \times \sin(\angle_B)$

        \end{center}

\addcontentsline{toc}{subsection}{\bf A.2: List of f6 Pipeline Files}{\bf \large A.2: List of f6 Pipeline Files}
        \label{sec: List of f6 Pipeline Files}   
        List of file names and types that are used in the f6* pipeline towards the production of f6u maps.
        
            \begin{table} [H]
\label{tab:product_table}
    \hspace*{-1.5cm}
    \begin{tabular}[ht]{llclp{10cm}}
        \hline
            File Name              & File      & New Data   & Internal or     & File Description                                                          \\ 
                                   & Format    & Product    & External        &                                                                           \\ \hline
            f61                    & .dat      &            & Int.            & Phase Correlation 1 $\times$ 3 skip                                       \\ 
            f62                    & .dat      &            & Int.            & Phase Correlation 2 $\times$ 2 skip                                       \\ 
            f63                    & .dat      &            & Int.            & Phase Correlation 3 $\times$ 2 skip                                       \\ 
            f6c                    & .txt      &            & Int.            & Duty cycle                                                                \\ 
            f6e                    & .fits.gz  & Yes        & Ext.            & 48 Hour fully calibrated Line of Site (LoS) phase shift map               \\ 
            f6f                    & .lst      &            & Int.            & List of fqi files for f6m map                                             \\ 
            f6g                    & .jpg      & Yes        & Ext.            & 48 Hour fully calibrated map for visualization                            \\ 
            f6g                    & .jpg      & Yes        & Ext.            & 48 Hour fully calibrated magnetic field strength map                      \\ 
            f6j                    & .jpg      & Yes        & Int.            & 24 Hour phase shift maps for visualization                                \\ 
            f6j                    & .jpg      & Yes        & Ext.            & 24 Hour phase shift maps for visualization                                \\ 
            f6m                    & .fits     &            & Ext.            & Component phase shift maps                                                \\ 
            f6q                    & .fits.gz  & Yes        & Ext.            & 48 Hour fully calibrated Line of Site (LoS) magnetic field strength map   \\ 
            f6r                    & .jpg      & Yes        & Ext.            & Line of Sight for fully calibrated magnetic field strength maps           \\ 
            f6r                    & .jpg      & Yes        & Ext.            & 48 Hour fully calibrated Line of Site (LoS) magnetic field strength map   \\ 
            f6s                    & .jpg      & Yes        & Int.            & Line of Sight for 24 Hour phase shift maps                                \\ 
            f6s                    & .jpg      & Yes        & Ext.            & Line of Site for 24 Hour phase shift maps for visualization               \\ 
            f6u*                   & .txt      &            & Int.            & Legacy format ROI list                                                    \\ 
            f6u*\_dev              & .csv      & Yes        & Int.            & New ROI list used for development                                         \\  
            f6u*\_fs               & .fits.gz  & Yes        & Int.            & 200 $\times$ 500 composite phase shift map (Long/Sine(Lat))               \\ 
            f6u*\_fs\_global       & .jpg      & Yes        & Ext.            & Global phase shift map                                                    \\ 
            f6u*\_fs\_nolab        & .jpg      &            & Int.            & 200 $\times$ 500 phase shift map (no labels in Long/Sine(Lat))            \\ 
            f6u*\_fsrs             & .fits.gz  & Yes        & Ext.            & 200 $\times$ 500 composite phase shift map (Long/Lat)                     \\ 
            f6u*\_fsrs\_nolab      & .jpg      & Yes        & Int.            & 360 $\times$ 180 phase shift map (no labels in Long/Lat)                  \\ 
            f6u*\_fsrs\_problabel  & .jpg      & Yes        & Ext             & 360 $\times$ 180 phase shift map (probability labels in Long/Lat)         \\ 
            f6u*\_maplist          & .csv      & Yes        & Int.            & List of all potential 24 hour phase shift maps                            \\ 
            f6u*\_mask             & .fits.gz  &            & Int.            & 200 $\times$ 500 DOI mask (Long - Sine(Lat))                              \\ 
            f6u*\_maskrs           & .fits.gz  & Yes        & Int.            & 360 $\times$ 180 DOI mask (Long - Lat)                                    \\ 
            f6u*\_mg               & .fits.gz  &            & Int.            & 200 $\times$ 500 composite magnetic strength map (Long/Sine(Lat))         \\ 
            f6u*\_mg\_global       & .jpg      & Yes        & Ext.            & Global magnetic strength map                                              \\ 
            f6u*\_mg\_nolab        & .jpg      &            & Ext.            & 500 $\times$ 200 magnetic strength map (no labels in Long/Sine(Lat))      \\ 
            f6u*\_mgrs             & .fits.gz  & Yes        & Ext.            & 360 $\times$ 180 composite magnetic strength map (Long/Lat)               \\ 
            f6u*\_mgrs\_nolab      & .jpg      & Yes        & Ext.            & 360 $\times$ 180 magnetic strength map (no labels in Long/Lat)            \\ 
            f6u*\_mgrs\_problabel  & .jpg      & Yes        & Int.            & 360 $\times$ 180 magnetic strength map (probability labels in Long/Lat)   \\ 
            f6u*\_ns               & .fits.gz  &            & Int.            & 200 $\times$ 500 nearside magnetogram (Long/Sine(Lat))                    \\ 
            f6u*\_nsrs             & .fits.gz  & Yes        & Int.            & 360 $\times$ 180 nearside magnetogram (Long/Lat)                          \\ 
            f6x                    & .txt      & Yes        & Ext.            & Table showing all detected active regions with probability>50\%           \\ 
            f6z                    & .fits     &            & Int.            & Line-of-sight phase shift maps                                            \\ 
            \hline \\ 
        \end{tabular}
    \end{table}
    
       \clearpage    

\end{document}